\documentclass[aps,prr,floatfix,showpacs,twocolumn,longbibliography,superscriptaddress]{revtex4-1}
\usepackage{ulem}
\usepackage{dcolumn}
\usepackage{bm}
\usepackage{color}
\usepackage{amssymb}
\usepackage{amsmath}
\usepackage{graphicx}
\usepackage{amsfonts}
\usepackage{slashed}
\usepackage{pstricks}
\usepackage{float}
\usepackage{multirow}
\usepackage{array}
\allowdisplaybreaks
\usepackage{dcolumn}
\usepackage{epsf}
\usepackage{float}
\usepackage[nice]{nicefrac}

\begin{document}
\title{Hidden-nucleons neural-network quantum states for the nuclear many-body problem}

\author{Alessandro Lovato}
\affiliation{Physics Division, Argonne National Laboratory, Argonne, IL 60439}
\affiliation{INFN-TIFPA Trento Institute of Fundamental Physics and Applications, 38123 Trento, Italy}

\author{Corey Adams}
\affiliation{Physics Division, Argonne National Laboratory, Argonne, IL 60439}
\affiliation{Leadership Computing Facility, Argonne National Laboratory, Argonne, IL 60439}

\author{Giuseppe Carleo}
\affiliation{Institute of Physics, École Polytechnique Fédérale de Lausanne (EPFL), CH-1015 Lausanne, Switzerland}

\author{Noemi Rocco}
\affiliation{Theoretical Physics Department, Fermi National Accelerator Laboratory, P.O. Box 500, Batavia, Illinois 60510, USA}

\date{\today}
\begin{abstract}
We generalize the hidden-fermion family of neural network quantum states to encompass both continuous and discrete degrees of freedom and solve the nuclear many-body Schr\"odinger equation in a systematically improvable fashion. We demonstrate that adding hidden nucleons to the original Hilbert space considerably augments the expressivity of the neural-network architecture compared to the Slater-Jastrow ansatz. The benefits of explicitly encoding in the wave function point symmetries such as parity and time-reversal are also discussed. Leveraging on improved optimization methods and sampling techniques, the hidden-nucleon ansatz achieves an accuracy comparable to the numerically-exact hyperspherical harmonic method in light nuclei and to the auxiliary field diffusion Monte Carlo in $^{16}$O. Thanks to its polynomial scaling with the number of nucleons, this method opens the way to highly-accurate quantum Monte Carlo studies of medium-mass nuclei.

\end{abstract} 

\maketitle

\section{Introduction}
Since the early 2000s, the combination of nuclear effective field theories (EFTs) and sophisticated many-body methods has paved the way to a systematic description of atomic nuclei that is well rooted in the underlying theory of strong interactions~\cite{Hergert:2020bxy}. In particular, numerical methods based on single-particle basis expansions, such as self-consistent Green’s function~\cite{Soma:2020xhv}, the in-medium similarity-renormalization group~\cite{Hergert:2015awm}, and Coupled Cluster~\cite{Hagen:2013nca} have achieved a formidable success in treating binding energies and radii up to $^{208}$Pb~\cite{Hu:2021trw} and have provided a plausible solution to the quenching puzzle of beta-decays~\cite{Gysbers:2019uyb}. Nevertheless, some prominent challenges remain open. As a chief example, most of existing nuclear Hamiltonians cannot simultaneously describe light nuclear systems and the equation of state of infinite nucleonic matter~\cite{Nosyk:2021pxb,Lovato:2022apd}. Exhaustive tests of nuclear interactions require many-body methods that are suitable to retain the complexity of nuclear dynamics at short distances~\cite{Piarulli:2019pfq,Benhar:2021doc}. An accurate description of the latter is also critical to reproduce exclusive lepton-nucleus scattering cross sections~\cite{CLAS:2018xvc,Pybus:2020itv,CLAS:2020mom}, for the calculation neutrinoless double-beta decay matrix elements~\cite{Weiss:2021rig}, and for studying neutron-star matter in the high-density regime~\cite{Piarulli:2019pfq,Lovato:2022apd,Benhar:2022yme}.

Continuum quantum Monte Carlo (QMC) methods~\cite{Carlson:2014vla} have no difficulties in treating short-range (or high-momentum) components of the nuclear wave function, but are presently limited to light nuclei, with up to $A=12$ nucleons. The primary reason preventing QMC calculations of medium-mass and large nuclei is the onset of the fermion-sign problem. Controlling it requires employing sophisticated wave-functions, whose calculation either scales exponentially with the number of nucleons~\cite{Pudliner:1995wk,Pudliner:1997ck} or violates the factorization theorem~\cite{Lonardoni:2017hgs,Lonardoni:2018nob,Novario:2021low}. Therefore, extending QMC calculations to medium-mass nuclei rests on our ability to develop wave functions that capture the vast majority of nuclear correlations while requiring computational time that scales polynomially with $A$. Developing such wave-functions is also important to correctly evaluate observables that do not commute with the Hamiltonian --- e.g. spatial and momentum distributions, electroweak transition and responses --- as their QMC estimates strongly depend upon the quality of the wave-function ansatz. 

After the earliest application to prototypical interacting spins models~\cite{Carleo:2017} artificial neural networks (ANNs) have proven to compactly and accurately represent the wave function of a variety of strongly interacting systems~\cite{Choo:2019,Hermann:2019,Pfau:2019}. The applicability of ANNs in solving the nuclear Schr\"odinger equation in momentum space has been first demonstrated in Ref.~\cite{Keeble:2019bkv}, where the ground-state of the deuteron is reproduced with remarkable accuracy by a single-layer ANN --- see Ref.~\cite{Sarmiento:2022bxn} for a detailed uncertainty-quantification analysis. Subsequently, an anti-symmetric coordinate-space ansatz defined through the product between a permutation-invariant ANNs Jastrow and a Slater determinant of single-particle orbitals has been utilized in a variational Monte Carlo (VMC) method to solve leading-order pionless-EFT Hamiltonians of $A\leq 6$ nuclei~\cite{Adams:2020aax,Gnech:2021wfn}. Detailed comparisons against Green's function Monte Carlo and the hypershperical harmonics approaches have validated the expressivity of this ANN Slater-Jastrow (ANN-SJ) ansatz, but also highlighted its shortcomings, essentially due to the incorrect nodal surface of the Slater determinant. 

The authors of Ref.~\cite{Javier:2021} have recently introduced an extremely expressive family of variational wave functions, consisting of augmented Slater determinants involving ``hidden'' additional fermionic degrees of freedom. This family of wave functions, which has proven to be universal for fermionic lattice systems, has been utilized to study the Hubbard model on the square lattice. In this paper, we generalize this approach to encompass both discrete and continuous degrees of freedom and solve the nuclear many-body problem in a systematically-improvable fashion. We couple the ``hidden nucleons'' wave function with the VMC method developed in Refs.~\cite{Adams:2020aax,Gnech:2021wfn} to compute the ground-state energies of $^3$H, $^3$He, $^4$He, and $^{16}$O nuclei starting from the pionless-EFT Hamiltonian of Ref.~\cite{Schiavilla:2021dun}. We note that $^{16}$O becomes the largest nucleus studied with neural-network quantum states so far. Both parity and time-reversal symmetries are directly encoded in the variational state to appreciably improve the optimization of the ANN's parameters. The training is further accelerated capitalizing on a novel version of the stochastic reconfiguration~\cite{Sorella:2005} method, with a regularization based on RMSProp algorithm~\cite{Hinton:2018}. Our results for $^3$H, $^3$He, $^4$He nuclei are benchmarked against the numerically-exact hyperspherical-harmonics (HH) method~\cite{Kievsky:2008es}. For the larger $^{16}$O nucleus, we compare the hidden-nucleon ground-state energies and point-nucleon density with the ones computed with the AFDMC. 

This work is organized as follows. Section~\ref{sec:method} provides an overview of the nuclear Hamiltonian, the hidden-nucleon wave function, and of the optimization algorithm. In Section~\ref{sec:results} we discuss our results, comparing them with existing hyperspherical-harmonics and QMC nuclear methods. Finally, in Section~\ref{sec:conclusions} we state our conclusions and provide perspectives on future developments.

\section{Methods} 
\label{sec:method}
To a remarkably large extent, the dynamics of atomic nuclei can be modeled by nonrelativistic Hamiltonians of the form
\begin{equation}
H = - \sum_{i} \frac{\nabla_i^2}{2m_N} + \sum_{i<j}v_{ij} + \sum_{i<j<k} V_{ijk}\, ,
\label{eq:hamiltonian}
\end{equation}
where $v_{ij}$ and $V_{ijk}$ denote the nucleon-nucleon ($NN$) and three-nucleon $3N$ potentials. In this work, following  Ref.~\cite{Adams:2020aax,Gnech:2021wfn}, we employ $NN$ and $3N$ forces derived at leading order in a pionless-EFT expansion that consist of contact terms between nucleons~\cite{Bedaque:2002mn}. More specifically, we use the leading-order $NN$ potential ``o'' of Ref.~\cite{Schiavilla:2021dun} that is designed to reproduce the $np$ scattering lengths and effective radii in $S/T = 0/1$ and $1/0$ channels. We assume the electromagnetic component of the $NN$ potential to only include the Coulomb repulsion between finite-size protons. The authors of Ref.~\cite{Schiavilla:2021dun} explored different regulator values for the $3N$ force. Here we take $R_3= 1.0$ fm because when used in conjunction with model ``o'',  this value of $R_3$ provides binding energies that are in reasonably good agreement with experiments for various closed-shell nuclei across the nuclear chart. 

To fix the notation, we introduce $R=\{\mathbf{r}_1\dots \mathbf{r}_A\}$ and $S=\{\mathbf{s}_1\dots \mathbf{s}_A\}$ to indicate the set of single-particle spatial three-dimensional coordinates and the z-projection of the spin-isospin degrees of freedom $\mathbf{s}_i = \{s^z_i , t^z_i \}$ of the $A$ nucleons comprising a given nucleus. 

\subsection{Hidden-nucleon wave function}

Generalizing the ``hidden-fermion'' approach of Ref.~\cite{Javier:2021} to encompass continuous and discrete coordinates, we introduce a ``hidden-nucleon'' (HN) ansatz to model the ground-state wave functions of atomic nuclei. In addition to the visible coordinates $R,S$, the Hilbert space also contains fictitious degrees of freedom for the $A_h$ hidden nucleons, which are in turn functions of the visible ones $(R_h,S_h) = f(R,S)$. The amplitudes of the hidden-nucleons wave function ansatz in the $R,S$ basis are schematically given by
\begin{equation}
\Psi_{HN}(R,S)\equiv \det
\begin{bmatrix}
    \phi_v(R,S) & \phi_v(R_h,S_h)\\
    \chi_h(R,S) & \chi_h(R_h,S_h)\\
  \end{bmatrix}
\end{equation}
In the above equation, $\phi_v(R,S)$ denotes the $A \times A$ matrix representing visible single-particle orbitals computed on the visible coordinates --- this would be the only component of the wave function in a Hartree-Fock description of the nucleus. Note that, in contrast with Ref.~\cite{Javier:2021}, the columns of our matrix denote different particles, while rows refer to different states. The $A_h \times A_h$ matrix $\chi_h(R_h,S_h)$ yields the amplitudes of hidden orbitals evaluated on hidden coordinates. Finally, $\chi_h(R,S)$ and $\phi_v(R_h,S_h)$ are $A_h \times A$ and $A \times A_h$ matrices that provide the amplitudes of hidden orbitals on visible coordinates and those of visible orbitals on hidden coordinates, respectively. 

In the limit $A_h=1$, $\chi_h(R,S)=0$, and $\chi_h(f(R,S))=0$ we recover the usual Slater-Jastrow formulation, which can be thereby interpreted as a limit of the hidden-nucleon ansatz. If the function $f$ is permutation invariant, it is immediate to prove that $\Psi_{HN}(R,S)$ is anti-symmetric under the exchange of two-particles coordinates. The authors of Ref.~\cite{Javier:2021} have demonstrated that the hidden-fermion ansatz can represent any anti-symmetric functions on the lattice, provided that the functions $\chi_h$ and $f$ are general.

In order to bypass the combinatorial nature of the function $f$, we directly parametrize each column ``$i$'' of $\phi_v(R_h,S_h)$ and $\chi_h(R_h,S_h)$ in terms of independent, real-valued, permutation-invariant neural networks as 
\begin{align}
\phi_v^i(R_h,S_h) &= e^{\mathcal{U}^i_\phi(R,S)}\tanh[\mathcal{V}^i_\phi(R,S)]\nonumber\\
\chi_h^i(R_h,S_h) & = e^{\mathcal{U}^i_\chi(R,S)}\tanh[\mathcal{V}^i_\chi(R,S)]\,.
\end{align} 
As in Refs.~\cite{Adams:2020aax,Gnech:2021wfn,Pescia:2021kxb}, permutation-invariance is achieved by expressing the functions $\mathcal{U}^i_\phi$, $\mathcal{V}^i_\phi$, $\mathcal{U}^i_\chi$, and $\mathcal{V}^i_\chi$ in terms of the Deep-Sets architectures~\cite{Zaheer:2017,Wagstaff:2019}. Taking pair coordinates as input instead of single-particle ones was found to accelerate the training when the sum pooling is used~\cite{Gnech:2021wfn}. Here, we achieve similar performances with single-particle inputs employing the \textit{logsumexp} pooling
\begin{equation}
    \mathcal{F}(R,S)= \rho_\mathcal{F}\Big[\log\Big(\sum_i e^{\phi_\mathcal{F}(\mathbf{r}_i,\mathbf{s}_i)}\Big)\Big]
\end{equation}
Both $\phi_\mathcal{F}$ and $\rho_\mathcal{F}$ are dense feed-forward neural networks, comprised of two hidden layers with $16$ nodes each. The latent space, i.e. the size of the output of $\phi_\mathcal{F}$ and the input of $\rho_\mathcal{F}$, is $16$-dimensional. The output layers of $\rho_\mathcal{F}$ contain $A$ nodes for $\mathcal{F} =\mathcal{U}^i_{\phi}$ and $\mathcal{F} =\mathcal{V}^i_{\phi}$ and $A_h$ nodes for  $\mathcal{F} =\mathcal{U}^i_{\chi}$ and $\mathcal{F} =\mathcal{V}^i_{\chi}$. 

The single-particle orbitals defining $\phi_v(R,S)$ and $\chi_h(R,S)$ are represented by dense feed-forward neural networks that take as input the single particle coordinates of the nucleons. They are both comprised of two hidden layers with $10$ nodes each, while their outputs are one-dimensional with linear activation functions.

The hyperbolic tangent is the activation function chosen for the hidden layers of all the dense feed-forward neural networks used in this work. We tried alternative differentiable functions, such as \textit{Softplus}~\cite{Dugas:2000} and \textit{GELU}~\cite{Hendrycks:2016}, without finding appreciable differences. Note that our choices are restricted to differentiable activation functions because the calculation of the kinetic energy requires evaluating their second derivatives. 

\subsection{Symmetries and sampling}

In this work, we enforce point-symmetries, such as parity and time reversal, into the hidden nucleon ansatz. Since all the atomic nuclei considered in this work have ground-states with positive parity, we can construct such variational states by 
\begin{align}
\Psi_{HN}^{P}(R,S) \equiv \Psi_{HN}(R,S) + \Psi_{HN}(-R,S)\,. 
\label{eq:psi_P}
\end{align}
For even-even nuclei, such as $^4$He and $^{16}$O, we can additionally enforce time-reversal symmetry 
\begin{align}
\Psi_{HN}^{PT}(R,S) \equiv \Psi_{HN}^P(R,S) + \Psi^P_{HN}(R,\theta S)\,,
\label{eq:psi_PT}
\end{align}
where $\theta S$ is obtained by applying the operator $-i\sigma_y$ to all single-particle spinors~\cite{bohr1998nuclear}. Note that no complex conjugate operation is required in the above definition since $\Psi_{HN}^P(R,S)$ is a real-valued function. As discussed in Sec.~\ref{sec:results}, enforcing point-symmetries considerably accelerates the training and augments the expressivity of the ANN for a fixed number of hyperparameters. In principle, additional point-symmetries, such as ``signature'', and ``simplex''~\cite{Dobaczewski:1999gs}, can be imposed in a similar fashion as Eqs.~\ref{eq:psi_P} and ~\ref{eq:psi_PT}. Their impact on ANN quantum states will be investigated in future works.  

The set of variational parameters $\mathbf{p}$ entering the ANN variational state are optimized minimizing the expectation value of the Hamiltonian $E = \langle \Psi | H | \Psi\rangle /  \langle \Psi | \Psi\rangle$. The $3A$ spatial dimensional integrals and the spin-isospin summations are evaluated in a stochastic fashion, using the Metropolis-Hastings sampling algorithm discussed in the supplemental material of Ref.~\cite{Adams:2020aax}. It has to be noted that the hidden-nucleon ansatz is completely general and does not prevent the Metropolis-Hastings algorithm to sample nonphysical states. To better elucidate this issue, consider a nucleus with a given total isospin projection $T_z$. Even though nuclear interactions are charge conserving, the hidden-nucleon ansatz yields non-zero amplitudes for states $S$ such that $\sum_i t_i^z \neq T_z$. The latter are avoided by constraining the Metropolis walk onto states with $\sum_i t_z^i = T_z$, which is equivalent to multiplying $\Psi_{HN}(R,S)$ by the Kronecker Delta $\delta(\sum_i t_z^i - T_z)$. Similarly, the LO pionless-EFT Hamiltonian of Eq.~\ref{eq:hamiltonian} does not contain a tensor force and preserves the total spin projection on the z-axis $S_z$. As a consequence, to prevent sampling nonphysical states, we restrict the Metropolis walk to $S$ with $\sum_i s_i^z = S_z$. In practice, this is achieved by starting the walk with a state that satisfies the above requirements and sampling the new ones by exchanging the spin and the isospin projections of two, not necessarily different, pairs of nucleons. 

\subsection{Optimization}

Given the neural-network architecture, and the sampling procedure as described in the previous discussions, we further need to specify a procedure to minimize the variational energy and find the optimal parameters of the model wave function. The stochastic-reconfiguration (SR) algorithm~\cite{Sorella:2005}, closely related to the natural gradient descent method~\cite{amari_natural_1998} in unsupervised learning, has proven to efficiently optimize neural network quantum states for a variety of applications, ranging from spin models~\cite{Carleo:2017,bukov:2021,Javier:2021} to periodic bosonic systems~\cite{Pescia:2021kxb} and atomic nuclei~\cite{Adams:2020aax,Gnech:2021wfn}. The variational parameters are updated as $\mathbf{p}_t = \mathbf{p}_{t-1} -\eta S^{-1}_{t-1} \mathbf{g}_{t-1}$, where $\eta$ is the learning rate, $S$ is the quantum Fisher-information matrix, and $\mathbf{g} = \partial E/\partial\mathbf{p}$ is the gradient of the energy. The inversion of matrix $S$ is typically stabilized by adding a small positive diagonal matrix $S \to S + \epsilon I$, implying that all diagonal elements are shifted by the same amount, thereby neglecting potential order of magnitudes differences in the parameters' changes~\cite{Sorella:2007}. To remedy this shortcoming, inspired by the RMSProp method~\cite{Hinton:2018}, we accumulate the exponentially-decaying averages of the squared gradients 
\begin{equation}
\mathbf{v}_t = \beta \mathbf{v}_{t-1} + (1 - \beta) \mathbf{g}_t^2
\end{equation}
and regularize the the Fisher-information matrix by $S \to S + \epsilon \, {\rm diag} (\sqrt{\mathbf{v}_t} + 10^{-8})$. In the original formulation of the SR algorithm, taking larger values of $\epsilon$ reduces the magnitude of the parameters' update and rotates it towards the stochastic gradient descent direction. In our RMSProp version, the regularization term rotates the update towards the RMSProp direction, which typically yields faster training than the simple stochastic gradient descent. Since the dimension of the Fisher matrix scales quadratically with the number of variational parameters, storing it in memory becomes unfeasible for a nucleus as large as $^{16}$O. To overcome this limitation, as in Refs.~\cite{Neuscamman:2012}, we use the iterative conjugate-gradient method to solve the linear system associated with the SR parameters' update. 
Finally, the {\it Adaptive Epsilon} heuristic scheduler introduced in Ref.~\cite{Adams:2020aax} has been adopted to determine the best value of the regularization parameter $\epsilon$ at each optimization step.

\begin{figure}[t]
\begin{center}
	\includegraphics[width=0.49\textwidth]{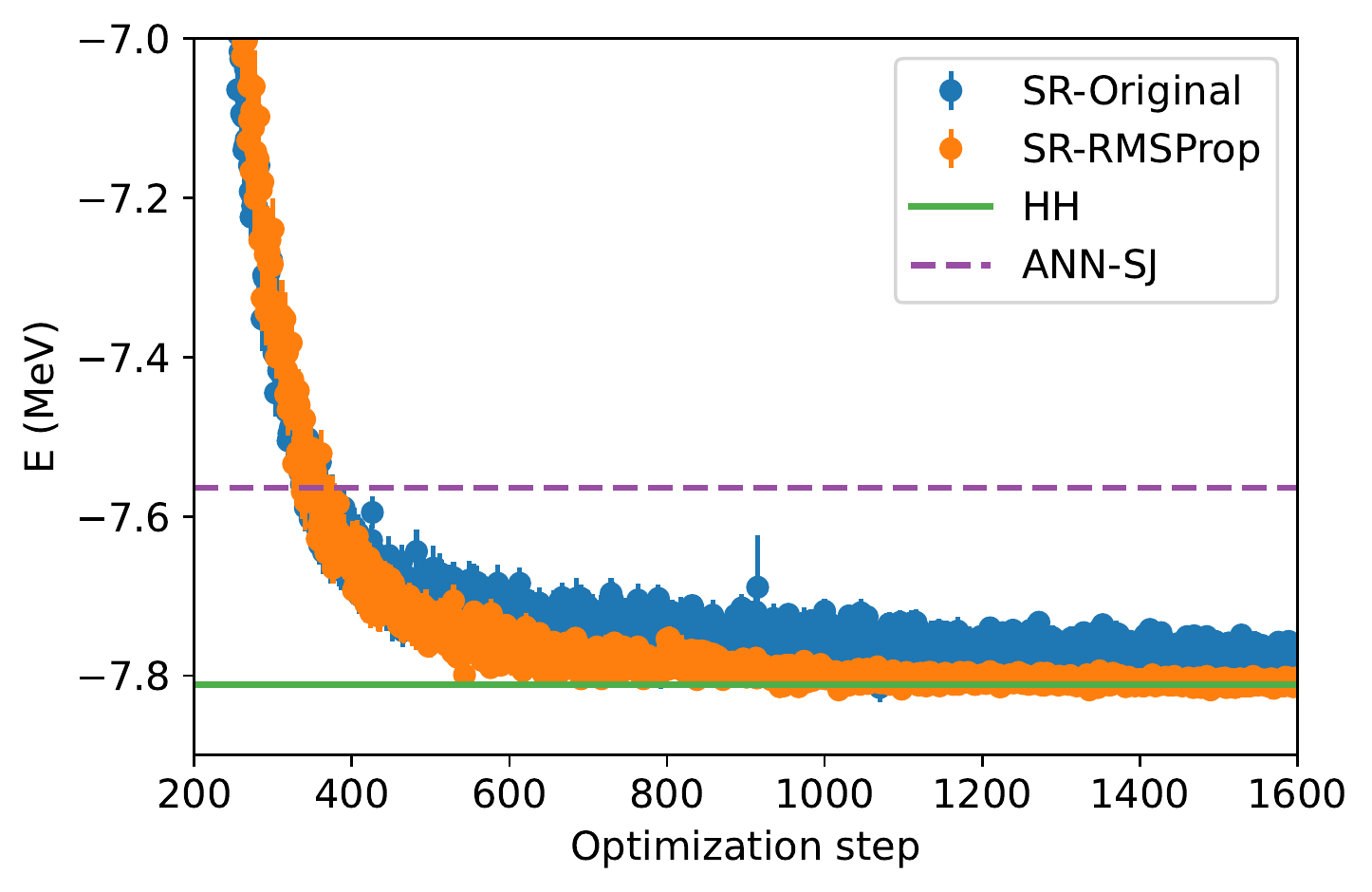} 
	\includegraphics[width=0.49\textwidth]{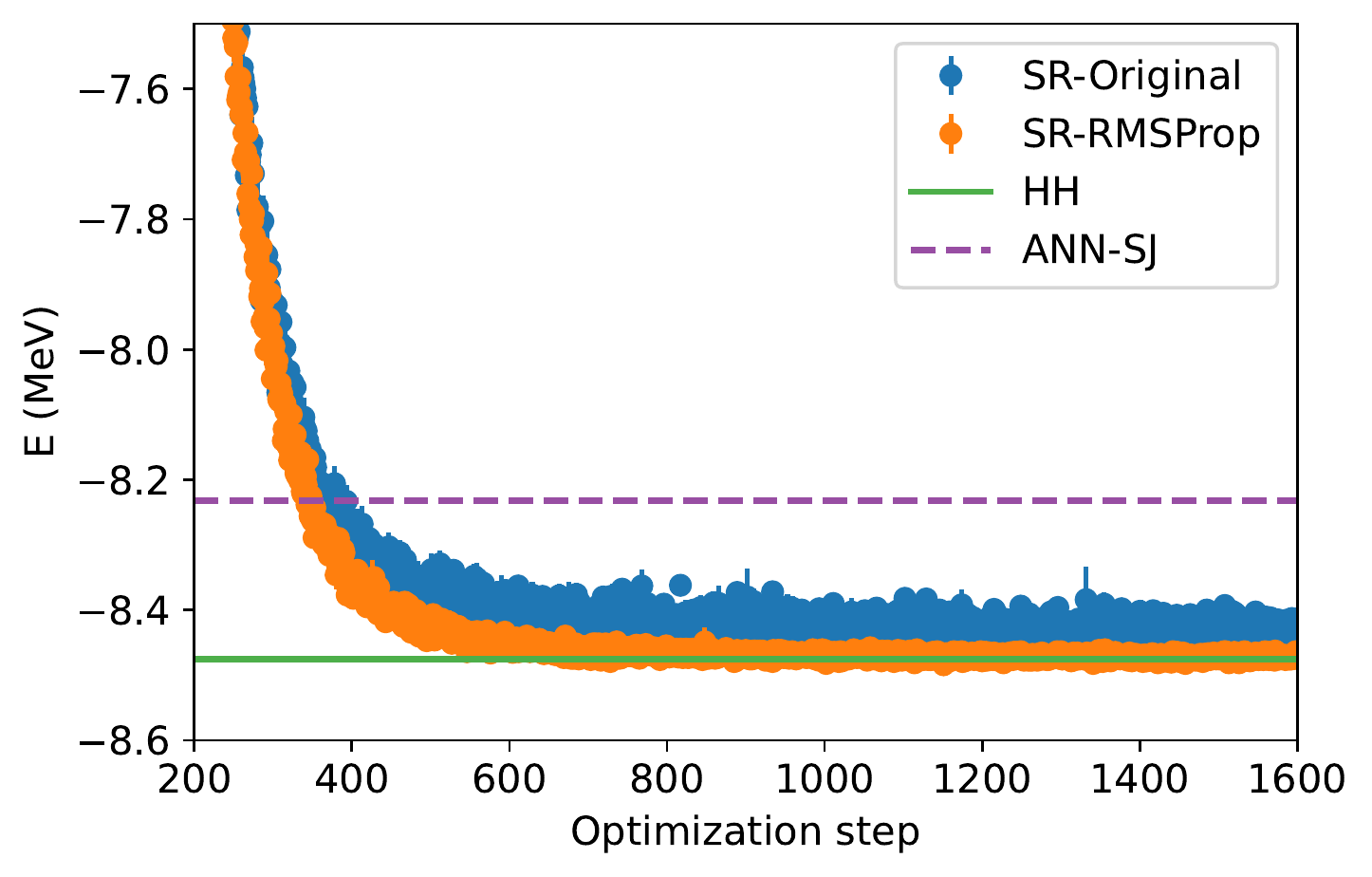} 
	\caption{Convergence of the SR algorithm for $^3$He (upper panel) and $^3$H (lower panel) with the original (blue solid circles) and RMSProp-like (orange solid circles) diagonal shifts. The ANN-SJ and the HH energies of Ref.~\cite{Gnech:2021wfn} are displayed by the purple dashed and solid green lines, respectively.}
	\label{fig:3He}
\end{center}
\end{figure}

\begin{figure}[b]
\begin{center}
	\includegraphics[width=0.49\textwidth]{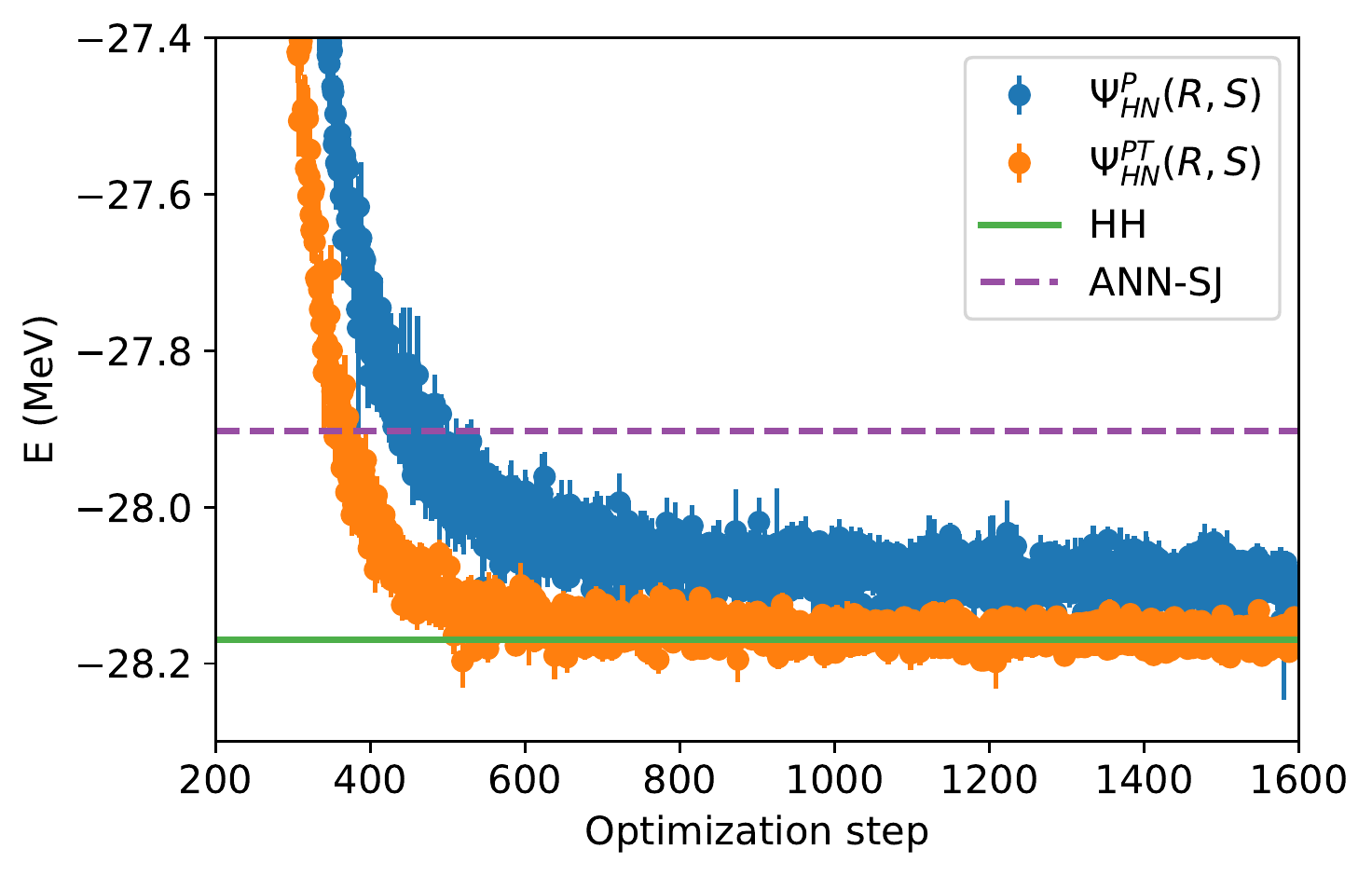} 
	\includegraphics[width=0.49\textwidth]{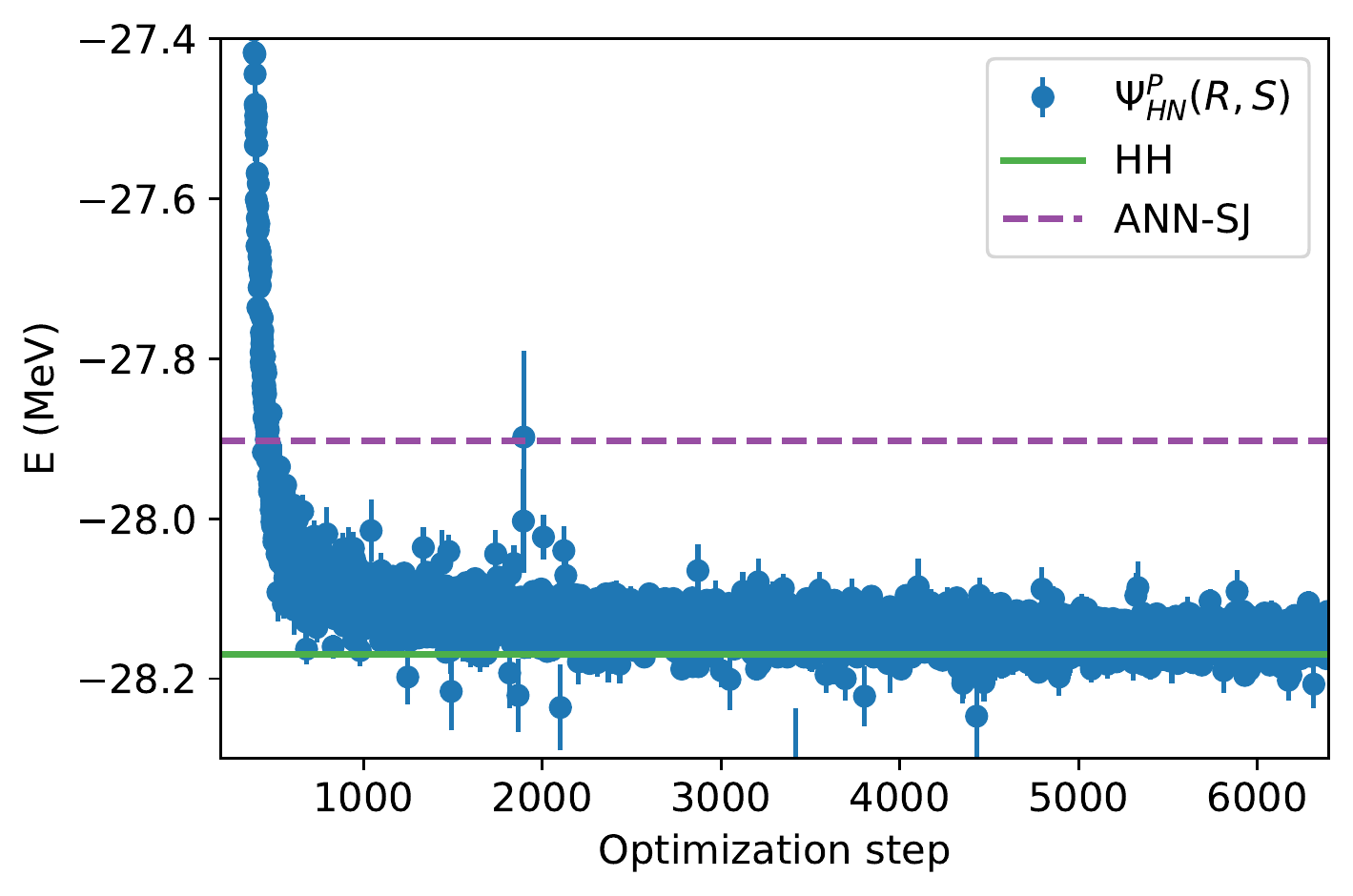} 
	\caption{Upper panel: $^4$He ground-state energy convergence with the ansatz that conserves parity (blue solid circles) and  parity plus time reversal (orange solid circles). The ANN-SJ and the HH ground-state energies of Ref.~\cite{Gnech:2021wfn} are displayed by the purple dashed and solid green lines, respectively.
	Lower panel: Convergence of the parity-conserving ansatz utilizing a wider ANN than in the upper panel.}
	\label{fig:4He}
\end{center}
\end{figure}

\section{Results} 
\label{sec:results}

We begin our analysis comparing the performances of the original SR method with the version that includes the RMSProp-inspired diagonal shift in the Fisher-information matrix. In Fig.~\ref{fig:3He}, we show the convergence of the ground-state energy of $^3$He (upper panel) and $^3$H (lower panel) as obtained with $A_h=3$ hidden nucleons and the positive-parity ansatz of Eq.~\eqref{eq:psi_P}.  The blue solid circles, corresponding to the energies obtained using the RMSProp-like regularization, are noticeably closer to the numerically-exact HH result of Ref.~\cite{Adams:2020aax} than those obtained with the original version of the SR algorithm. The fact that the SR-RMSProp estimates are less scattered that the SR ones is another indication of the better minima found by new version of the algorithm. Most notably, independent of the particular regularization choice, both the SR and SR-RMSProp energies are appreciably lower than the ANN-SJ value reported in Ref.~\cite{Gnech:2021wfn}. 

Because of its superior training performances with respect to the original version of the algorithm, in the remainder of the paper we will only show results obtained with the SR-RMSProp minimizer. The convergence of the $^4$He ground-state energy computed with $A_h=4$ hidden nucleons is in displayed in Fig.~\ref{fig:4He}. The parity-conserving wave function $\Psi_{HN}^{P}(R,S)$ is outperformed by $\Psi_{HN}^{PT}(R,S)$, which additionally preserves the time-reversal symmetry. Both of them provide significantly better energies than the original ANN-SJ model, as they can improve the nodal surface of the single-particle Slater determinant. More importantly, $\Psi_{HN}^{PT}(R,S)$ provides a variational energy that is consistent with the numerically-exact HH estimate of Ref.~\cite{Gnech:2021wfn}. It has to be noted that $\Psi_{HN}^{P}(R,S)$ should in principle converge to the exact energy, even with $A_h=4$ hidden nucleons, but it requires wider (or deeper) ANN architectures. To prove this point, in the lower panel of Fig.~\ref{fig:4He} we show the training of $\Psi_{HN}^{P}(R,S)$ with $A_h=4$ in which the number of nodes in the hidden layers in $\phi_\mathcal{F}$ and $\rho_\mathcal{F}$ has been increased from $16$ to $24$. After about $4800$ optimization steps, the parity-conserving ansatz yields energies that are consistent with the HH method. Nevertheless, our results indicate that enforcing time-reversal symmetry is effective in reducing the training time and augment the expressivity of the hidden-nucleon ANN architecture. 

Neural-network quantum states applications to nuclear systems have so far been limited to light nuclei, with up to $A=6$ nucleons~\cite{Keeble:2019bkv,Adams:2020aax,Gnech:2021wfn}. Here, we significantly extend the reach of this methods by computing the ground-state of $^{16}$O utilizing the hidden-nucleon ansatz. In Ref.~\cite{Schiavilla:2021dun}, the AFDMC method has been employed to study this nucleus using as input the LO pionless-EFT Hamiltonian of Eq.~\eqref{eq:hamiltonian}. The AFDMC trial wave function takes the factorized form $\Psi_T( R, S ) \equiv \langle R S | \mathcal{F} |\Phi \rangle$. The Slater determinant of single-particle orbitals $\Phi( R, S )$ determines the long-range behavior of the wave function. The correlation operator is expressed as 
\begin{equation}
\mathcal{F} = \Big( \prod_{i<j<k} F^c_{ijk} \Big) \Big(\prod_{i<j} F^c_{ij}\Big) \Big (1 + \sum_{i<j} F^{op}_{ij} \Big) 
\label{eq:afdmc_corr}
\end{equation}
The spin-isospin independent three-body correlations $F^c_{ijk}$ act on all triplets of nucleons. Similarly, the central two-body Jastrow $F^c_{ij}$ is applied to all nucleon pairs, while the spin-isospin dependent term, $F^{op}_{ij}$, appears in a linearized form~\cite{Gandolfi:2014ewa}. This approximation reduces the computational cost of evaluating $\Psi_T( R, S )$ from exponential to polynomial in $A$ but makes the trial wave function non extensive: if the system is split in two (or more) subsets of particles that are separated from each other, the $\mathcal{F}$ does not factorize into a product of two factors in such a way that only particles belonging to the same subset are correlated. As a consequence, the correlation operator of Eq.~\eqref{eq:afdmc_corr} becomes less effective for nuclei larger than $^{16}$O, preventing the applicability of the AFDMC method to medium-mass nuclei. 

\begin{figure}[t]
\begin{center}
	\includegraphics[width=0.49\textwidth]{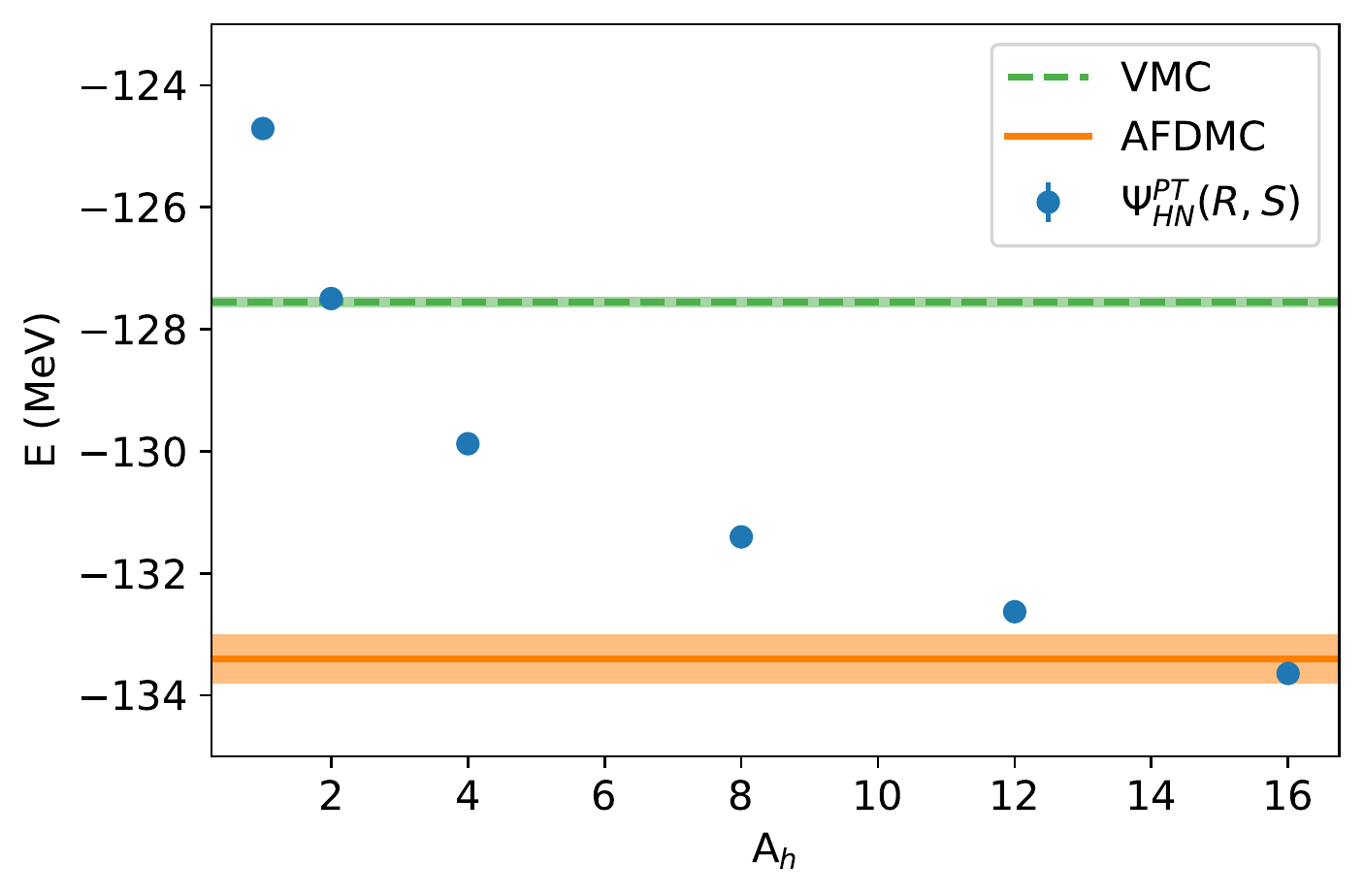} 
	\caption{Ground-state energy of $^{16}$O as a function of the number of hidden nucleons $A_h$ (solid blue points). The VMC and AFDMC energies --- the latter taken from Ref.~\cite{Schiavilla:2021dun} --- are shown by the green-dashed and orange solid lines. The shaded areas represent the Monte Carlo statistical uncertainties.}
	\label{fig:16O_energy}
\end{center}
\end{figure}

The AFDMC projects out the ground-state of the system from the starting trial wave function performing an evolution in imaginary time $\tau$
\begin{equation}
|\Psi_0 \rangle \propto \lim_{\tau\to\infty} |\Psi(\tau)\rangle = e^{-H \tau}|\Psi_T\rangle\, .
\end{equation} 
The fermion-sign problem is mitigated by means of the constrained-path approximation, which essentially limits the imaginary-time propagation to regions where the propagated and trial wave functions have a positive overlap~\cite{Carlson:2014vla}. Contrary to the fixed-node approximation, the constrained-path approximation does provide an upper bound to the true ground-state energy of the system~\cite{Wiringa:2000gb}. The accuracy of the trial wave function is critical to reduce this bias, as the constrained-path approximation becomes exact when the trial wave function is coincides with the ground-state one.  

In Fig.~\ref{fig:16O_energy}, we display the ground-state energy of $^{16}$O as a function of the number of hidden nucleons $A_h$ for the parity and time-reversal conserving ansatz of Eq.~\eqref{eq:psi_PT}. For comparison, the VMC energy of $^{16}$O obtained with the correlation operator of Eq.~\eqref{eq:afdmc_corr} is represented in Fig.~\ref{fig:16O_energy} by the dashed green line, while the shaded area is the Monte Carlo statistical uncertainty. The solid horizontal line and the shaded area indicate the constrained-path AFDMC energy and its statistical uncertainty as listed in Ref.~\cite{Schiavilla:2021dun}. Already for $A_h=2$, the hidden-nucleon wave function matches the VMC value. By further increasing $A_h$, the variational energy lowers until it becomes consistent with the AFDMC value, within error bars, demonstrating the accuracy of the hidden-nucleon ansatz even in the p-shell region. 

\begin{figure}[t]
\begin{center}
	\includegraphics[width=0.49\textwidth]{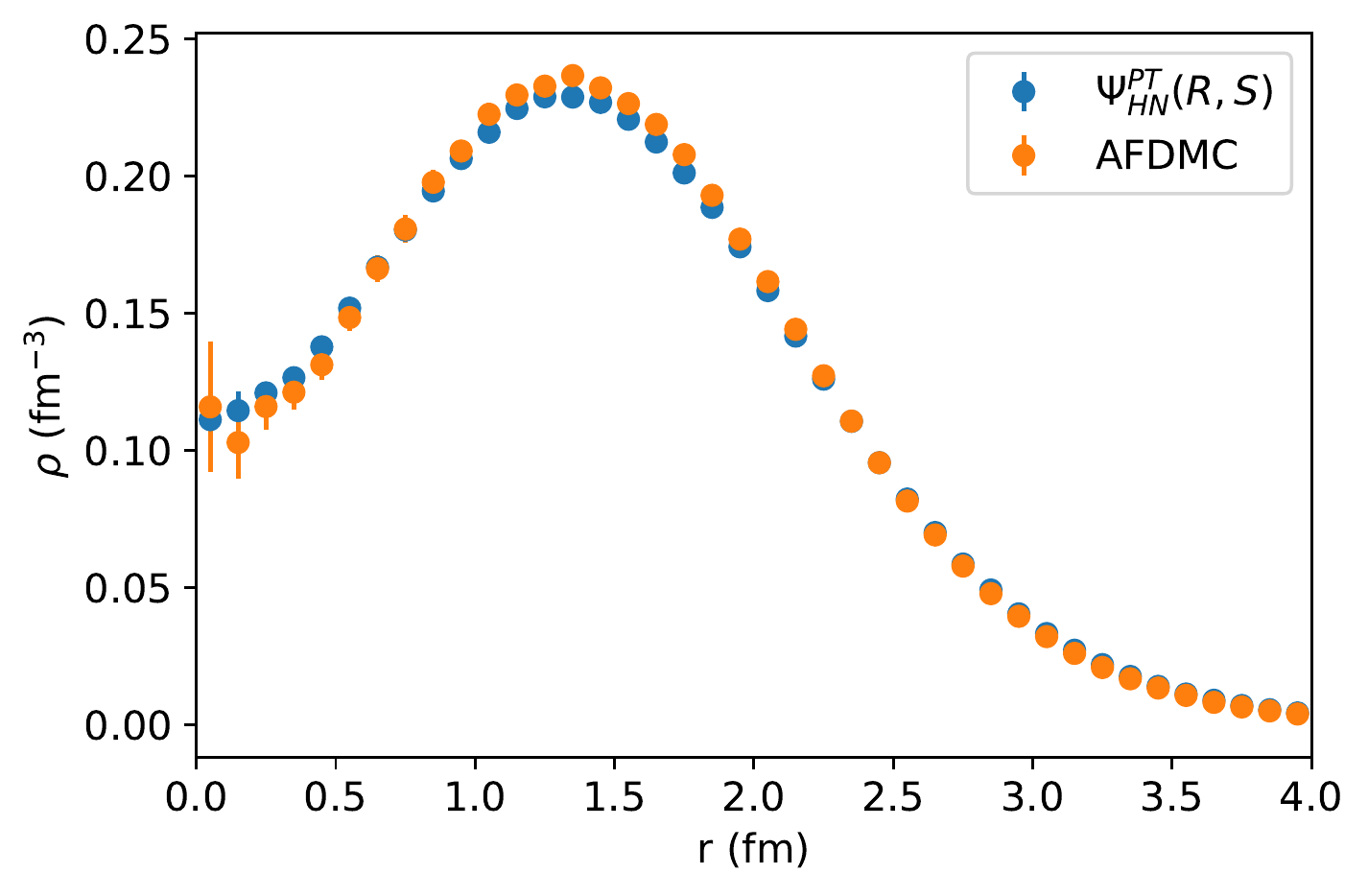} 
	\caption{Point nucleon density of $^{16}$O as obtained with the hidden nucleon ansatz (solid blue circles) compared with the perturbatively-corrected AFDMC estimates of Eq.~\eqref{eq:pc}.}
	\label{fig:16O_rho}
\end{center}
\end{figure}

Unless a forward-walk propagation is used~\cite{Zhang:1993,Casulleras:1995}, within diffusion Monte Carlo methods, expectation values of operators that do not commute with the Hamiltonian are typically estimated at first order in perturbation theory as 
\begin{equation}
\frac{\langle\Psi(\tau) | O | \Psi(\tau)\rangle}{\langle\Psi(\tau) | \Psi(\tau)\rangle} \simeq 
2 \frac{\langle\Psi_T | O | \Psi(\tau)\rangle}{\langle\Psi_T | \Psi(\tau)\rangle} - \frac{\langle\Psi_T | O | \Psi_T\rangle}{\langle\Psi_T | \Psi_T\rangle}\, .
\label{eq:pc}
\end{equation}
Hence, in addition to controlling the fermion-sign problem, the accuracy of $\Psi_T( R, S )$ is critical to evaluate observables that do not commute with the Hamiltonian, such as density distributions in coordinate and momentum space. In any case, the extrapolated estimator is always biased in a quantity difficult to assess. On the other hand, neural-network quantum states provide pure estimators and no extrapolations are required for computing expectations values of operator that do not commute with the Hamiltonian. 

In Fig.~\ref{fig:16O_rho}, we display the point-nucleon density of $^{16}$O as obtained with $A_h=16$ hidden nucleons, compared with the AFDMC results, which are obtained in perturbation theory as in Eq.~\eqref{eq:pc}. Remarkably, despite the lack of mean-field information encoded in the hidden nucleon anstaz --- the single-particle orbitals are randomly-initialized feed-forward neural networks --- the training procedure yields a point-nucleon density of $^{16}$O that is very close to the AFDMC one. This excellent agreement confirms the accuracy of the hidden-nucleon ansatz in modeling the wave functions of atomic nuclei at both short and long distances.

\section{Conclusions} 
\label{sec:conclusions}
We have developed a novel, hidden-nucleon, neural-network ansatz suitable to solve the nuclear many-body Schr\"odinger equation of a leading-order pionless-EFT Hamiltonian in a systematically improvable fashion. To this aim, we extend the hidden-fermion family of variational states to encompass both continuous and discrete degrees of freedom, corresponding to the spatial and spin-isospin coordinates of the nucleons. Point symmetries such as parity and time reversal are built in the neural-network wave function to augment its expressivity. We have concurrently improved the stochastic-reconfiguration algorithm by introducing a non-constant diagonal regularization inspired by the RMSProp optimization method. 

We first gauge the hidden-nucleon wave function in $^3$H, $^3$He, and $^4$He nuclei, whose ground-state energies turn out in excellent agreement with numerically exact hyperspherical-hamonics results reported in Ref.~\cite{Gnech:2021wfn}. In $A=3$ nuclei, we observe that the RMSProp-inspired diagonal shift considerably accelerates the convergence of the training compared to the default version of the SR method and adopt it for all the subsequent calculations. We observe that enforcing time-reversal symmetry noticeably accelerates the optimization of the $^4$He ground-state wave function and improves the expressivity of the neural-network ansatz for a fixed number of variational parameters.  
   
We then applied the hidden-nucleon architecture to study the ground-state of $^{16}$O, the largest nuclear system 
yet computed with neural-network quantum states. Even using only two hidden nucleons, we are able to match the VMC energies obtained with conventional spin-isospin dependent variational wave-functions. By further increasing $A_h$, the hidden-nucleon architecture gains the capability
to reproduce the AFDMC constrained-path energy, thereby proving the expressivity of this ansatz for continuous degrees of freedom. 

In addition to the ground-state energy, the neural-network architecture is capable of learning the single-particle density of $^{16}$O, at both short and long distances. This is particularly remarkable, as, in contrast with other QMC methods, the hidden-nucleon ansatz is agnostic to the mean-field properties of the specific nucleus of interest. Besides the Hamiltonian of choice, the only information entering the calculation are the number of protons and neutrons, the total spin, and its point symmetries; the short- and long-range components of the ground-state wave functions are learned by minimizing the energy expectation value.

The hidden-nucleon ansatz provides a compact representation of nuclear wave function in terms of a relatively-small number of variational parameters. For instance, the ground-state wave function of $^{16}$O with $A_h$ hidden nucleons is completely determined by about 77,000 neural-network parameters. This is an order of magnitudes smaller than the number of coefficients required to represent the wave functions by state-of-the-art nuclear many-body methods based on a single-particle basis expansion, such as the no-core shell model~\cite{Barrett:2013nh}. The wave functions computed in this work, as well as the Google-JAX program used to generate them are publicly available~\cite{JAX-QMC}. They can be readily used to compute quantities of experimental interests, without the need of training the ANN.  

In principle, generalized backflow transformations similar to those employed to model the ground-state of electronic systems~\cite{Hermann:2019, Pfau:2019} can be used to augment the expressivity of the hidden-nucleon architecture. We anticipate using them when dealing with more sophisticated nuclear forces, characterized by strong tensor components. Since the latter do not conserve the total spin projection on the z-axis $S_z$, it may well be beneficial to employ an over-complete spin basis, similar to the employed by the AFDMC method.

Finally, it has to be noted that the hidden-nucleon approach exhibit a favorable polynomial scaling in computational time with the number of nucleons. The calculation of the wave function requires $A^3$ operations, while evaluating the expectation value of the spin-isospin dependent component of the NN potential costs $A^5$ operations. On the other hand, including spin-isospin dependent 3N forces without resorting to density-dependent approximations would make the method to scale as $A^6$. We envision treating medium-mass nuclei as large as $^{40}$Ca by exploiting leadership class hybrid GPU/CPU computers.

\section{Acknowledgments}
Enlightening discussions with K. E. Schmidt and F. Vicentini are gratefully acknowledged. The present research is supported by the U.S. Department of Energy, Office of Science, Office of Nuclear Physics, under contracts DE-AC02-06CH11357, by the NUCLEI SciDAC program (A.L.), the DOE Early Career Research Program, and Argonne LDRD awards, and by Fermi Research Alliance, LLC under Contract No. DE-AC02-07CH11359 with the U.S. Department of Energy, Office of Science, Office of High Energy Physics (N.R.). This research used resources of the Argonne Leadership Computing Facility, which is a DOE Office of Science User Facility supported under Contract DE-AC02-06CH11357 and the Laboratory Computing Resource Center of Argonne National Laboratory. Part of the numerical calculations presented in this work were carried out through a CINECA-INFN agreement that provides access to resources on MARCONI100 at CINECA.

\bibliography{biblio}

\end{document}